\documentstyle{article}
\newcommand{\be}{\begin{equation}}
\newcommand{\ee}{\end{equation}}
\def\bea{\begin{eqnarray}}
\def\eea{\end{eqnarray}}

\def\b{\beta}

\def\f{\phi}

\def\l{\lambda}
\def\m{\mu}
\def\n{\nu}

\def\r{\rho}

\def\t{\tau}

\def\D{\Delta}

\def\L{\Lambda}

\def\S{\Sigma}

\begin{document}

\bibliographystyle{unsrt}
\footskip 1.0cm
\thispagestyle{empty}
\setcounter{page}{0}
\begin{flushright}
DFTUZ/96/11\\
Imperial/TP/95-96/34\\
March 1996\\
\end{flushright}
\vspace{10mm}

\centerline {\LARGE SYMMETRY NON RESTORATION AND}
\vspace{5mm}
\centerline {\LARGE INVERSE SYMMETRY BREAKING ON THE LATTICE}
\vspace{5mm}
\centerline {\LARGE}
\vspace*{15mm}
\centerline {\large G. Bimonte }
\centerline {\it  Departamento de F\'{\i}sica Te\'orica, Facultad de
Ciencias}
\centerline {\it Universidad de Zaragoza, 
50009 Zaragoza, SPAIN}
\vspace*{5mm}
\centerline{ \small and }
\vspace*{5mm}
\centerline{ \large G. Lozano \footnote{
E-mail addresses: bimonte@napoli.infn.it~~,~~g.lozano@ic.ac.uk}}
\centerline{\it  Imperial College, Theoretical Physics}
\centerline{\it Prince Consort Road, SW7 2BZ London, ENGLAND}
\vspace*{15mm}
\normalsize

\centerline{\bf Abstract}
\vspace{10mm}

{\large We study the  finite temperature symmetry behaviour of
$O(N_1) \times O(N_2)$ scalar models on the lattice and we 
prove that at sufficiently high temperatures and in arbitrary 
dimensions their full symmetry
is always restored or, equivalently, that the phenomenon 
of Symmetry Non Restoration which, according to lowest order
perturbation theory, takes place in the continuum version of 
these models, does not occur on the lattice.}

\newpage

\baselineskip=24pt
\setcounter{page}{1}


The high temperature behaviour of relativistic field theories has 
been the subject of an 
intense research 
since the early works by Kirzhnitz and Linde \cite{lin}, Weinberg 
\cite{wei} and Dolan and Jackiw \cite{dol}. The 
results of these investigations, show that
in a "typical" case the symmetry of the vacuum increases when the
temperature is raised and thus, in spontaneously broken theories,
internal symmetries are gradually restored when the system is
heated up.
Thus, in the context of Grand Unified Theories and Cosmology
the fact that the universe might have been in phases with different 
symmetry properties 
at different stages of its evolution,
undergoing
a series of phase 
transitions in the process of cooling 
down, 
would have significant consequences, like 
for instance the creation of topological defects 
via the Kibble
mechanism \cite{kib}.

Nevertheless, as noticed by Weinberg \cite{wei},  
$O(N_1) \times O(N_2)$
theories
may present an ``atypical" symmetry behaviour, where  by atypical, 
we mean that either the symmetry is not restored at high temperatures 
or that 
an exact symmetry 
of the low temperature theory becomes broken at higher temperatures.
These phenomena, known as Symmetry Non Restoration (SNR) and 
Inverse Symmetry
Breaking (ISN) are in fact two aspects of the same problem, the only
difference 
between 
them being whether or not the symmetry is 
broken in the zero
temperature theory. 
To {\it lowest order in perturbation theory}, the 
existence  of SNR or ISB is related to the possibility 
of having negative
Debye masses. This can be achieved in multi-scalar theories, 
provided that
some of the fourth-order couplings are taken negative and large 
enough in absolute value (but also small enough as to 
produce a bounded potential).
As on the other hand, the scalar sector
of most extensions of the Standard Model and Grand Unified 
Theories is
rather undetermined, it turns out that SNR and ISB are not 
so atypical
as one would first suppose.
 
Indeed, it has been recognised that the ideas of SNR and ISB 
can have interesting phenomenological implications. Very 
recently for instance,
the phenomenon of SNR has been used by 
Dvali, Melfo and Senjanovi\'{c}
\cite{goran} to suggest 
that the monopole problem might not exist in some GUT's, by arguing that
the monopole-producing phase transition might have never occurred 
(for an earlier implementation of this idea see 
\cite{sal}).
Also in connection to  monopoles, ISB is the basis of
the proposal by Langacker and Pi \cite{la},
which states that a period of broken $U(1)_{em}$ 
can
 cure the monopole problem. Other no less interesting applications
of SNR and ISB concern the breaking of the CP symmetry \cite{ms},
the domain wall problem \cite{dvali}, 
Baryogenesis \cite{dod},
Inflation \cite{lee} and P, Strong CP, and Peccei-Quinn Symmetries 
\cite{dva2}. ISB and SNR have also been considered in 
 \cite{kuz}.

Appealing as the idea of SNR and ISB may be,
 no consensus has yet been reached on whether they really
correspond to a true physical effect or rather to an artifact of
( may be, lowest order?) perturbation theory. This second point 
of view stems
from the fact that when non-perturbative approximations are used
to study the symmetry behaviour of these models
\cite{fuj}, it is found that
symmetry is invariably restored at high temperature.
Moreover, even staying within the realms of perturbation theory, it has
been shown \cite{bim} that the inclusion of next to leading 
order effects in the 
calculation of thermal masses tends to reduce the region of parameter
space in which SNR and ISB occur. Recently \cite{ame}, 
the gap equations used in
\cite{bim} have been rederived by a more detailed analysis based in the
Cornwall-Jackiw-Tamboulis effective potential \cite{corn} at 
finete temperature \cite{amepi}.
An independent study, which also
encodes 
some non perturbative information through the 
Effective Action Technique \cite{wet}, has
been carried out
\cite{roos}. 
For the small values of the scalar self-couplings which were 
considered in \cite{roos}, it was found 
that the corrections to the lowest order
perturbative computation are small. While this is not in
disagreement with the results of \cite{bim}, which also
predict small corrections when these couplings are small, we 
think that an extension of the analysis of \cite{roos} to include large
couplings may be unavoidable, since, as pointed out in 
\cite{goran}, realistic models may require large scalar self-couplings,
due to the presence of gauge interactions which
conspire against SNR and ISB. 

Due to the conflicting results which emerge from the 
perturbative, semi perturbative and non perturbative methods 
we mentioned above, we
think that Lattice Field Theory might 
result to be a useful setting
to study SNR and ISB.  
Some work along this lines has been already done in \cite{wip} where
it is shown that, when an approximation  based on  the constraint 
effective potential \cite{ora} is made, symmetry is always restored in less 
than four dimensions.

In this letter, we will study SNR and ISB on the Lattice without making
approximations and in arbitrary dimensions and we will prove that
symmetry is always restored, at sufficiently high temperatures,  for 
$O(N_1) \times O(N_2)$ models. In doing this, we will closely follow
the steps of a theorem by King and Yaffe on symmetry restoration for $O(N)$
models on the Lattice.
It is worth mentioning 
that while King and Yaffe's result
for $O(N)$  models confirms the result obtained from perturbation theory, the
result we present here exactly contradicts the 
lowest order perturbative calculation.

The continuum model we inicially consider is a
global $O(N_1)\times O(N_2)$
scalar theory in $(d+1)$ euclidean dimensions described by the action,
\be
S=\int d^{d+1}x\left\{\sum_{i=1,2}\left[\frac{1}{2}\mid \partial_{\m}
\f_i\mid^2+\frac{1}{2}m^2_i \mid \f_i \mid^2
+ \l_i(\mid \f_i \mid^2)^2\right]-\l 
\mid \f_1\mid^2\mid \f_2\mid^2\right\}~,\label{cont}
\ee
where $\f_i\equiv (\f_i^{(1)},\cdots , \f_i^{(N_i)})$
 is an $N_i$-component real scalar and
$\mid \f_i \mid^2=\sum_{j=1}^{N_i} (\f_i^{(j)})^2$. The
condition of boundedness for the potential constrains the
coupling constants to satisfy the relations:
\be
\l_i>0~,~~~~~4\l_1 \l_2> \l^2~.
\ee
This condition allows for positive values of $\l$ and, as it immediately
follows from
the one-loop computation \cite{wei},
if $\l$ is such that,
\be
\l > \frac{4+2N_2}{N_1}\l_2
\ee
the $O(N_1)\times O(N_2)$ symmetry is necessarily
broken to $O(N_1)\times O(N_2-1)$ at high $T$. The symmetry
at lower temperatures depends instead on the signs and 
magnitudes of the masses $m^2_i$.

We will now 
prove  that when the
model (\ref{cont}) is defined on a
discrete lattice of points, 
the full $O(N_1)\times O(N_2)$
symmetry {\it is restored} at sufficiently high
temperatures, for all values of the parameters in the
action.

We consider an anisotropic hypercubic $(d+1)$-dimensional
lattice 
$\L\equiv {\cal T}\times \S$. Here, $\S$ is an infinite
$d$-dimensional hypercubic lattice accounting for space, while
${\cal T}$ is a finite one-dimensional lattice consisting of
$N_{\t}$ points, accounting for the finite euclidean time
axis. We assume a priori distinct spacings
$\D x$ and $\D \t$ for $\S$ and $\cal T$ respectively.
Thus:
$$
\L \equiv {\cal T}\times \S
=\{x=(x_0,~\bar{x}) :~x_0=n_0 \D \t~~~~
\bar{x}=(n_1 \D x,\cdots,n_d \D x)~ ;
$$
\be
n_0=1,\cdots,N_{\t};~~~~~
n_i\in{\bf Z}~~~~i=1,\cdots,d\}~~.   \label{lat}
\ee
The temperature of the system is then given by:
\be
T=\frac{1}{N_{\t}\D \t}~.
\ee 
The lattice version of the model is described by the action:
$$
S=\sum_{x \in \L}\D \t (\D x)^d\left\{
\sum_i\left[\frac{1}{2}\left(
\frac{1}{(\D \t)^2}
\mid 
\hat{\f}_i(x+e_{\t}\D \t)-
\hat{\f}_i(x)\mid ^2+
\sum_{k=1}^d\frac{1}{(\D x)^2}
\mid\hat{\f}_i(x+e_{k}\D x)-\hat{\f}_i(x)\mid ^2 \right)\right.
\right.
$$
\be
+\left. \left. \frac{1}{2}\hat m_i^2 \mid \hat{\f}_i(x)\mid^2
+ \hat{\l}_i (\mid \hat{\f}_i(x)\mid^2)^2\right]
-\hat{\l}\mid\hat{\f}_1(x)\mid^2\mid\hat{\f}_2(x)\mid^2\right\}~.
\label{latact}
\ee
where $e_{\t}$ and $e_k$ are unit vectors pointing along the time 
and $k$-th spatial directions respectively
and the  fields $\hat{\f}_i(x)$ satisfy periodic boundary
conditions in the time direction:
\be
\hat{\f}_i(x_0+\D \t N_{\t},\bar{x})=
\hat{\f}_i(x_0,\bar{x})~.\label{bcon}
\ee
It is convenient to introduce the dimensionless quantities:
$$
\hat{\f}_i(x)=(\D x)^{\frac{1-d}{2}}\f_i(x),~~a=\frac{\D \t}{\D x},~~
$$
\be
m_i^2=(\D x)^2\hat m_i^2,~~
\l_i=(\D x)^{3-d}\hat{\l}_i,~~\l=(\D x)^{3-d}\hat{\l}.\label{resc}
\ee
In terms of them the action (\ref{latact}) reads:
$$
S=
\sum_{x \in \L}\left\{
\sum_i\left[\frac{1}{2}\left(
\frac{1}{a}
\mid {\f}_i(x+e_{\t}\D \t)-{\f}_i(x)\mid ^2+
\sum_{k=1}^d a \mid {\f}_i(x+e_{k}\D x)-{\f}_i(x)\mid ^2 \right)\right.
\right.
$$
\be
+\left. \left. \frac{1}{2}a m_i^2 \mid {\f}_i(x)\mid^2
+ a{\l}_i (\mid {\f}_i(x)\mid^2)^2\right]
-a{\l}\mid {\f}_1(x)\mid^2\mid {\f}_2(x)\mid^2\right\}~
\label{latact2}
\ee

It is also convenient to measure the temperature in units of
the inverse of the lattice spacing $\D x$. We thus define
the ``lattice temperature''
\be
T_L=(\D x) T=\frac{1}{N_{\t}a}\equiv \frac{1}{\b_L}~.
\ee

A way to test the symmetry of the system is to consider 
the two-point functions:
\be
\langle 
\overline{\f_j(z)}\cdot \f_j(w)\rangle
\equiv Z^{-1}\int \left(
\prod_{x \in \L} 
\prod_{i=1,2} 
d \f_i(x)\right)\,
\overline{\f_j(z)}\cdot \f_j(w)\, \exp(-S)~,\label{corr}
\ee
where $Z$ is the partition function and $z$ and $w$ are two
spacelike separated points of $\L$. In the broken phases,
one or both of the correlators (\ref{corr}) have a non-vanishing
limit $M^2_j$ for infinite separations:
\be
M^2_j=\lim_{
\mid z - w\mid 
\rightarrow \infty}
\langle 
\overline{\f_j(z)}\cdot \f_j(w)\rangle \neq 0~~~~{\rm for~some~j}
\label{th}
\ee
while in the symmetric phase both these limits vanish. Now, we
will prove that, for all values of the mass parameters and
coupling constants in (\ref{latact2}),
and in any dimension $d$,
there exists a temperature $T_L^*$ above which the correlators
(\ref{corr}) decay exponentially with the separation
$\mid z - w\mid $:
\be
\mid \langle \overline{\f_j(z)}\cdot \f_j(w)\rangle \mid
\leq c \sqrt{\frac{T_L}{\n}}\, \exp[{-M(T_L)\mid z - w\mid}]~~~~T_L
\geq T_L^*~.
\label{bound}
\ee
In this equation $c$ is a numerical constant, 
$M(T_L)$ is a function such that:
\be
M(T_L)>0~~~{\rm for}~~ T_L>T_L^*~;~~~~~\lim_{T_L \rightarrow \infty}
M(T_L)=\infty
\ee
and $\n$ is a constant independent on $T_L$.
Once (\ref{bound}) is proved, it then will follow that above $T_L^*$ 
both limits $M^2_j$ vanish
and the full $O(N_1)\times O(N_2)$ symmetry is restored.

Before proceeding further, a few comments are in order:

a) the temperature $T_L^*$ provides only an upper bound for the
true critical temperature $T_L^c$ of the symmetry-restoring
phase transition. As pointed out already in \cite{KY}, this
bound is not expected to have the correct dependence on the
bare coupling constants, for large values of the latter.
Consequently, based on this bound, no conclusions can be
drawn for the critical temperature in the continuum limit,
whenever this limit exists.

b) the bound we are going to derive, like 
the one in \cite{KY} for the $O(N)$ models, depends on $a$ only
through the temperature 
 $T_L=a N_{\t}$. This will allow us to take the
continuum limit in the time direction
$a\rightarrow 0,~~a N_{\t}\rightarrow {\rm const}$
in a straightforward way.

We now turn to the proof, which is, as we mentioned 
before, a simple extension
of King and Yaffe theorem on symmetry restoration for 
$O(N)$ models \cite{KY}.
The intuitive idea behind it is that, at high
temperature, the system can be thought of as a collection
of oscillators sitting on the sites of the spatial lattice
$\S$, such that no order is possible.
Guided by this idea, we will write the action, the 
partition function and
the correlators in a way which will turn to be useful to 
derive the bound
(\ref{bound})

We will start by rewriting the action 
(\ref{latact2})
as:
\be
S=\sum_{\bar{x}\in \S} 
S_{\bar x} 
- \sum_{l\in \S^*} V(l)~,\label{S}
\ee
where $\S^*$ is the set of links in $\S$ and
$$
S_{\bar x} =
\sum_{x_0 \in {\cal T}}
\sum_i\left[\frac{1}{2a}
\mid {\f}_i(x_0+\D \t,
\bar{x})-{\f}_i(x_0,\bar{x})\mid ^2+
\frac{1}{2}a (m_i^2+2d) \mid \f_i(x_0,\bar{x})\mid^2
+ a{\l}_i (\mid {\f}_i(x_0,\bar{x})\mid^2)^2\right]+
$$
\be
-\sum_{x_0 \in {\cal T}}
a{\l}\mid {\f}_1(x_0,\bar{x})\mid^2
\mid {\f}_2(x_0,\bar{x})\mid^2 ~,
\ee
while
\be
V(l)=\frac{1}{2}\sum_{x_0 \in {\cal T}}
\sum_{i=1,2}
a \mid {\f}_i(x_0,
\bar{x})+{\f}_i(x_0,\bar{y})\mid ^2~,\label{V}
\ee
where $\bar x$ and $\bar y$ are the end points of the link $l$. Now,
the first term on the r.h.s. of (\ref{S}) precisely describes a
set of uncoupled oscillators located at the sites of $\S$, while
the second sum provides an interaction among them. 
Notice also that the $V(l)$ is positive, a property which will be
important in what follows. 

Upon defining the measure
$$
d\m=\prod_{\bar{x}\in \S} 
d\m_{\bar x}
$$
where
\be
d\m_{\bar x}=z^{-1}\prod_{x_0 \in {\cal T}} 
\prod_{i=1,2}
 d \f_i (x_0, \bar{x})\,
\exp[-S_{\bar x}]~.\label{dmx}
\ee
and $z$ is a coefficient that normalises $d \mu_{\bar x}$ to one,
the partition function and the correlators can be written as
\be
Z=
\int d\m \,
\exp
\left[\sum_{l\in \S^*}V(l)\right]~,\label{Z}
\ee

\be
\langle 
\overline{\f_j(z)}\cdot \f_j(w)\rangle
\equiv Z^{-1}\int d \m \,\overline{\f_j(z)}\cdot \f_j(w)
\, \exp
\left[\sum_{l\in \S^*}V(l)\right]~.\label{c}
\ee

If we now define:
\be
\exp V(l)=1+\r(l)~,
\ee
and
\be
K(Y)=\int d\m \,\prod_{l\in Y}\r(l)~,
\ee
then
\be
Z=\int d\m
\,\prod_{l\in \S^*}[1+\r(l)]
=\sum_{Y \subset \S^*}
K(Y)~.\label{z2}
\ee

In the same way, defining
\be
K_j(Y)=\int d\m \,\overline{\f_j(z)}\cdot \f_j(w)
\prod_{l\in Y}\r(l)~,
\ee
we then have:
\be
\langle 
\overline{\f_j(z)}\cdot \f_j(w)\rangle
\equiv Z^{-1} \sum_{Y \subset \S^*}
K_j(Y)~.\label{c2}
\ee
Here
and in (\ref{z2}) the sum is over all subsets $Y$ of $\S^*$.

Let now $W$ 
be
the connected component of $Y$ that contains $w$ and
let $X=Y-W$. As the action $S_{\bar x}$ is invariant under
$\f_i \rightarrow -\f_i$, the only non-vanishing terms
in the sum (\ref{c2}) are those for which $z$ 
is in $W$.
Together with the fact that the measures $d\m_{\bar x}$ are
normalised to one, this allows us to write:
\be
\langle 
\overline{\f_j(z)}\cdot \f_j(w)\rangle=
\sum_
{\begin{array}{c}W \subset \S^*\\
z,w \in W\\
W {\it connected}\end{array}} K_j(W)
\sum_{X\subset \S^*-
\overline{W}}
\frac{1}{Z}K(X)
\ee
where $\overline W$ denotes the closure of $W$, namely the set of
links in $\S^*$ that share an end-point with some link in
$W$.

Up to here, we have only rewritten $Z$ and the correlation 
functions in a convenient way. We shall now start to look 
for bounds for these quantities. The positivity of $V(l)$ comes now into 
play in a crucial way,
as it immediately implies:
\be
\frac{1}{Z}
\sum_{X\subset 
\S^*-\overline{W}}K(X)=
\frac{\int d\m \,\exp
\left[\sum_{l\in\S^*-\overline{W}}V(l)\right]}
{\int d\m \,\exp
\left[\sum_{l\in \S^*}V(l)\right]}
\leq1~~~~~~~~~\forall W \subset \S^*~~,
\ee
which in turn implies the bound:
\be
\langle 
\overline{\f_j(z)}\cdot \f_j(w)\rangle \leq
\sum_{\begin{array}{c}W \subset \S^*\\
z,w \in W\\
W {\it connected}\end{array}} K_j(W)~.
\ee
The next step is
then to find
a bound for $K_j(W)$. First of all, we observe
that:
\be
\mid \r(l) \mid = \mid \exp V(l)-1\mid \leq \mid V(l)\mid
\exp V(l)~.
\ee
Moreover, by Schwarz's inequality, we have
\be
V(l)\leq 
\sum_{x_0 \in {\cal T}} 
\sum_{i=1,2}\left[
a\mid {\f}_i(x_0,
\bar{x})\mid ^2 +a\mid{\f}_i(x_0,\bar{y})\mid ^2\right]~,\label{Scw}
\ee
which implies the other bound
\be
\prod_{l\in W}V(l) \leq \sum_{\{q(\bar{x})\}}
\prod_{\bar{x} \in \S}
\left\{\sum_{x_0 \in {\cal T}} 
\sum_{i=1,2}
a\mid {\f}_i(x_0,
\bar{x})\mid ^2\right\}^{q(\bar{x})}~,
\ee
Here the sum 
is over all possible choices of the
non-negative integers $q(\bar{x})\leq 2d$ which vanish for all 
$\bar{x}\not\in W$ and are such that
\be
\sum_{\bar{x}\in\S}q(\bar{x})=\mid W\mid \label{conq}
\ee
where $\mid W \mid $ denotes the number of links in $W$.
Use of the inequality:
\be
\mid
\overline{\f_j(z)}\cdot \f_j(w)\mid\leq
\mid{\f_j(z)}\mid ^2+ \mid\f_j(w)\mid^2
\ee
finally implies the bound:
$$
\mid K_j(W)\mid \leq 
\sum_{\{q(\bar{x})\}}
\int
d\m  \,(\mid \f_j(z)\mid^2+\mid \f_j(w)\mid^2)
$$
\be
\times \prod_{\bar{x} \in \S}\left\{
\left[ \sum_{x_0 \in {\cal T}} 
\sum_{i=1,2}
a \mid \f_i(x_0,\bar{x})\mid^2\right]^{q(\bar{x})}
\exp \left[p(\bar{x})
\sum_{x_0 \in {\cal T}}
\sum_{i=1,2}
a \mid \f_i(x_0,\bar{x})\mid^2\right]\right\}
~. \label{bk}
\ee
The factor $p(\bar{x})$ in the above formula represents
the number of links in $W$ that have $\bar x$ as endpoint. 
Obviously
\be
p(\bar{x})\leq 2d ,~~~~~~~ \forall \bar{x}~.\label{p}
\ee

The important feature of eq.(\ref{bk}) is that $K_j(W)$ is
now bounded by a sum of products of independent one-dimensional integrals.
The next move is to bound the latter by Gaussian integrals.
To this purpose, we define:
\be
d\n_{\bar{x}}=z^{\prime -1} 
\prod_{x_0 \in {\cal T}} 
\prod_{i=1,2}
d \f_i(x_0,\bar{x})
\exp \left\{-\frac{1}{2}\left[\sum_{x_0 \in {\cal T}}
\sum_{i=1,2}
\frac{1}{a}
\mid 
\f_i(x_0+\D \t,\bar{x})-\f_i(x_0,\bar{x})
\mid ^2+ a \m^2 \mid \f_i(x_0,\bar{x})\mid^2
\right]\right\}~.\label{gauss}
\ee
In this equation, $z^{\prime}$ is a normalisation factor, while 
$\m^2$ represents a variational parameter, whose value will be
fixed at the end such as to provide the best bound. For 
simplicity, we have introduced a common 
parameter $\mu^2$ for both fields
$\f_i$, while in principle a better bound could be obtained by
letting distinct values. The derivation of the bound would in
this case be sightly more involved but 
the final bound
would not be qualitatively different from the one obtained
with one 
parameter only.

Let us now go back to (\ref{bk}): each term in the sum is, as
we said, a product of independent one-dimensional integrals, 
$I_{\bar x}$,
one for each $\bar{x} \in W$. There are now two cases
to be considered: whether $\bar x$ is distinct 
from $z$ and $w$
or not. In the first case, with the aid of (\ref{gauss}),
we can write the corresponding one-dimensional integral $I_{\bar x}$
as:
$$
I_{\bar {x}} = \frac{N_{\bar x}}{D_{\bar x}}
$$
with
\be
N_{\bar{x}}=
\int d\n_{\bar x} \left(
\sum_{x_0 \in {\cal T}} 
\sum_{i=1,2}
a \mid \f_i(x_0,\bar{x})\mid^2 \right) ^{q(\bar{x})}
\exp T_{\bar{x}}[p(\bar{x})]~,\label{N}
\ee
\be
D_{\bar {x}}=\int d\n_{\bar x} 
\exp T_{\bar{x}}[p=0] 
\label{D}
\ee
and
$$
T_{\bar{x}}[p(\bar{x})]=\sum_{x_0 \in {\cal T}}\sum_{i=1,2}
\left[\left( 
\frac{1}{2} \m^2-\frac{1}{2}
m_i^2+p(\bar{x})-2d\right)  a\mid \f_i(x_0, \bar{x})\mid^2 
-
a \l_i(\mid \f_i(x_0,\bar{x})\mid^2)^2\right]+
$$
\be
+\sum_{x_0 \in {\cal T}} 
a\l \mid\f_1(x_0,\bar{x})\mid^2\mid\f_2(x_0,\bar{x})\mid^2~.\label{T}
\ee
Now, it is trivial to check that $T_{\bar x}[p(\bar{x})]$
is bounded by:
\be
T_{\bar{x}}[p(\bar{x})]\leq
\frac{\b_L}{4 \l_1 \l_2-\l^2}
\left[\l_2\left(\frac{1}{2}\m^2-
\frac{1}{2}m_1^2
+p(\bar{x})-2d\right)^2 + 
\l_1\left(\frac{1}{2}\m^2-
\frac{1}{2}m_2^2
+p(\bar{x})-2d\right)^2 \right. +
$$
$$
\left.  + \l 
\left(\frac{1}{2}\m^2 -\frac{1}{2}m_1^2
+p(\bar{x})-2d\right) \left(\frac{1}{2}\m^2 -\frac{1}{2}m_2^2
+p(\bar{x})-2d\right) \right] 
\equiv
{\cal A}~,\label{a}
\ee
where we have used $\sum_{x_0\in {\cal T}}a=\b_L$. 
We then find
that:
\be
N_{\bar{x}} \leq \exp({\cal A})\int d\n_{\bar x} \, \left(
\sum_{x_0 \in {\cal T}} 
\sum_{i=1,2}
a\mid \f_i(x_0,\bar{x})\mid^2\right)
^{q(\bar{x})}~.\label{X}
\ee

To proceed further, we will need to use the following bounds
for the moments of a
Gaussian distribution:
\be
\int d\n_{\bar x} 
\mid\f_i(x_0^{(1)},\bar{x})\mid^2 
\cdots \mid\f_i(x_0^{(p)},\bar{x})\mid^2 \leq
(2p-1)!!\left[\frac{N_i}{\b_L}\left(\frac{1}{\m^2}+\frac{\b_L}{\m}
\right)\right]^p
\ee
$$
\int d\n_{\bar{x}}
\mid \f_i(x_0,\bar{x})\mid^2\geq \frac{N_i}{\b_L \m^2}~~,
$$
\be
\int d\n_{\bar{x}}
 \mid \f_1(x_0,\bar{x})\mid^2 \mid \f_2(x_0,\bar{x})\mid^2
\geq \frac{N_1N_2}{\b_L^2 \m^4}~~,
\ee
to write:
$$
N_{\bar{x}} \,\leq
\exp({\cal A})
\sum_{r=0}^{q(\bar{x})}
\left( \begin{array}{c} r \\ q(\bar{x})  \end{array} \right) 
\int d\n_{\bar{x}}
\left(\sum_
{x_0 \in {\cal T}}
 a \mid \f_1(x_0,\bar{x})\mid^2\right)^{r}\left(\sum_{x_0 \in {\cal T}}
 a \mid \f_2(x_0,\bar{x})\mid^2\right)^{q(\bar{x})-r}
$$
$$
\leq 
(4d-1)!! \sum_{r=0}^{q(\bar{x})}
\left( \begin{array}{c} r \\ q(\bar{x}) \end{array} \right) 
\left[N_1 \left( \frac{1}{\m^2}+\frac{\b_L}{\m}\right)\right]^r
\left[N_2 \left( \frac{1}{\m^2}+
\frac{\b_L}{\m}\right)\right]^{q(\bar{x})-r}
\exp{\cal A}=
$$
\be
=
(4d-1)!!
\left[ (N_1+N_2) \left(
 \frac{1}{\m^2}+\frac{\b_L}{\m}\right)
\right]^{q(\bar{x})}\exp{\cal A}~.\label{Y}
\ee
As for $D_{\bar x}$, we can use Jensen's inequality (see \cite{roe})
to bound it as:

$$
D_{\bar x} \geq  
\exp 
\int d\n_{\bar{x}} T_{\bar{x}}[0]
\geq
$$
\be
\exp \left\{ \sum_{i=1,2} \left[\frac{N_i}{\m^2}
\left(
\frac{1}{2}\m^2-\frac{1}{2}m^2_i-2d\right)-3\l_i\b_L
\left(
\frac{N_i}{\b_L\m^2}+\frac{N_i}{\m}\right)^2\right]
+\frac{\l N_1 N_2}{\b_L \m^4}\right\}
\equiv \exp(- \cal{B})
~.\label{jen}
\ee
Using (\ref{jen}) together with (\ref{Y}) we obtain:
$$
I_{\bar x} \leq (4d-1)!! 
\left[(N_1+N_2)
\left(
\frac{1}{\m^2}+\frac{\b_L}{\m}\right)
\right]^{q(\bar{x})}
\times
\exp(\cal{A}+\cal{B})
$$

If instead $\bar x$ coincides with either $z$ or $w$ the previous
bound has to be multiplied by a factor of 
$$
(4d+1) \frac{N_j}{\b_L}
\left(
\frac{1}{\m^2}+\frac{\b_L}{\m}\right)~.
$$
Since now the only sites $\bar x$ for which $p(\bar{x})$ or
$q(\bar{x})$ are different from zero are those which are
endpoints of some link in $W$ and since there are less than
$2\mid W \mid$ such sites, we can use the previous bounds and 
eq.(\ref{conq})to say
\be
\mid K_j(W) \mid \leq 2(4d+1) \frac{N_j}{\b_L}
\left(
\frac{1}{\m^2}+\frac{\b_L}{\m}\right)
\sum_{\{q(\bar{x})\}}
\left[(N_1+N_2)
\left(
\frac{1}{\m^2}+\frac{\b_L}{\m}\right)
\right]^{\mid W \mid}[(4d -1)!!]^{2
\mid W \mid} \exp(2\mid W \mid ({\cal A}+{\cal B}))~.\label{W}
\ee
The number of terms in the
sum above is less than $(2d)^{2\mid W\mid}$ (see \cite{KY}
for a proof) and thus we have:
\be
\mid K_j(W) \mid \leq 2(4d+1) \frac{N_j}{\b_L}
\left(
\frac{1}{\m^2}+\frac{\b_L}{\m}\right)
\left[\hat{c}(N_1+N_2)
\left(
\frac{1}{\m^2}+\frac{\b_L}{\m}\right)
\right]^{\mid W \mid}
\exp(2\mid W \mid ({\cal A}+{\cal B}))~,\label{W2}
\ee
where $\hat{c}=4d^2 [(4d-1)!!]^2$. If we now take:
\be
\m^2=\sqrt{\frac{\n}{\b_L}}~,\label{mu}
\ee
where $\n$ is independent on $\b_L$, there will be a $\bar{\b}_L$
such that, for $\b_L < \bar{\b}_L$
\be
\m \b_L < 1~~~~~~~~~~\mu^2 \geq \mid m_i^2 \mid~~~i=1,2~.
\ee
These relations, together with (\ref{p}) then imply:
$$
\mid K_j(W)\mid \leq 4(4d+1)\frac{N_j}{\sqrt{\b_L\n}}
\left[2 \hat{c}(N_1+N_2)\sqrt{\frac{\b_L}{\n}}\right]^{\mid W \mid}
$$
$$
\times \exp \left\{2\mid W \mid \left[2d(N_1+N_2)\sqrt{\frac{\b_L}{\n}}
+ \n \frac{\l_1+\l_2+\l}{4 \l_1 \l_2 - \l^2} +
\frac{1}{\n}(12 \l_1 N_1^2 + 12\l_2 N_2^2 )\right]\right\}
$$
\be
\equiv c \frac{N_j}{\sqrt{\b_L\n}}\exp\left[-\mid W \mid M^{\prime}
(\b_L)\right]~.\label{fin}
\ee
It is clear that $\lim_{\b_L \rightarrow 0}
M^{\prime}(\b_L)=\infty$ and thus there exists a $\b_L^* \leq \bar{\b}_L$
such that, for $\b_L < \b_L^*$, $M^{\prime}(\b_L)>0$.
The steps to derive the final bound
(\ref{bound}) from eq.(\ref{fin}) are exactly 
the same as in \cite{KY} and we omit repeating the short proof here.
$(\b_L^*)^{-1}$
thus represent our bound for the true critical temperature of
the symmetry-restoring phase transition:
\be
T^c_L< \frac{1}{\b_L^*}~.
\ee
$\b_L^*$ depends of course on
$\n,d,\l_i,\l$ and $m_i^2$ and one might further
exploit the freedom in the choice of $\n$ in order to get the
best bound; as the analogue best bound for $\b_L^*$ in the
$O(N)$ case \cite{KY} does not have the right behaviour for large
values of the bare couplings and as we believe that the
same will occur in our case,
we will not explicitly show it here.

The above proof can be extended to the case when the
$O(N_1)\times O(N_2)$ symmetry is gauged (partly or completely),
along the lines of \cite{KY}, and we refer the reader
to that paper for the details.

We have  proved  that, in arbitrary dimensions and at 
sufficiently high temperatures, the full
$O(N_1) \times O(N_2)$ symmetry is restored , or equivalently,
that the phenomenon of Symmetry Non Restoration does not occur on the lattice.
Notice, though, that the implications of our proof regarding 
Inverse Symmetry Breaking are weaker, in the sense that we 
can not exclude the possibility
of this phenomenon taking place at intermediate temperatures. 
We can only state
that if an ISB phase transition occurs at a given temperature, then
the system will necessarily undergo a symmetry restoring phase transition
at higher temperatures.

Whether these results are relevant for the continuum, when the
continuum limit exists, depends on the behaviour of the
critical temperature of the symmetry-restoring phase
transition, when the lattice spacing $\D x$ is taken to zero.
Analogously to the $O(N)$ case, our bound for $T_c$ diverges
when the bare couplings become large and thus it is not 
possible to say, based on this bound, if $T_c$ remains finite
in the continuum limit. We would like to stress, once more,
that this behaviour of the bound is not peculiar to the
case examined here, but appears also in the $O(N)$ case in
\cite{KY}, for which one knows that $T_c$ has a finite continuum
limit. An answer to this question using Monte Carlo simulations
is under current investigation.

\section*{Acknowledgements}

G.L is partially supported by a grant from Fundaci\'on 
Antorchas (Argentina)
administered by the British Council. G.L thanks Prof. T.W.Kibble for
hospitality at the Blackett Laboratory Theory Group.




\baselineskip=20pt
\begin{thebibliography}{99}

\bibitem{lin} D.A. Kirzhnits and A.D. Linde, Phys. Lett. B42 (1972) 471.

\bibitem{wei} S. Weinberg, Phys. Rev. D9 (1974) 3357.

\bibitem{dol} L. Dolan and R. Jackiw, Phys. Rev. D9  (1974) 3320.

\bibitem{kib} T. W. Kibble, J. Phys. A9 (1976), 1987; Phys. Rep. 67
(1980), 183.

\bibitem{goran} 
G. Dvali, A. Melfo and G. Senjanovi\'{c}, 
Phys. Rev. Lett. 75 (1996) 4559.


\bibitem{sal} P. Salomonson, B.-S. Skagerstam and A. Stern, Phys. Lett.
B151 (1985) 243.

\bibitem{la} P. Langacker and S.-Y. Pi, Phys. Rev. Lett. 45 (1980), 1.

\bibitem{ms} R.N. Mohapatra and G. Senjanovi\'{c}, Phys. Lett. B89 (1979)
57; Phys. Rev. Lett.  42 (1979) 1651; Phys. Rev. D20 (1979)
3390.

\bibitem{dvali} G. Dvali and G. Senjanovi\'{c}, 
Phys. Rev. Lett. 74 (1995) 5178;

\bibitem{dod} S.Dodelson and L.Widrow, Mod. Phys. Lett A5 (1990) 1623; 
Phys. Rev. D42 (1990) 326; Phys. Rev. Lett.64 (1990) 340; 
S.Dodelson, B.R.Greene and L.M.Widrow, Nucl. Phys. B372 (1992) 347.

\bibitem{lee} J.Lee and I.Koh, 
``Inflation and inverse symmetry breaking", hep-ph/9506415.

\bibitem{dva2} G. Dvali, A. Melfo and G. Senjanovi\'{c}, 
``Non restoration of Spontaneously Broken P, CP and PQ
at High Temperature'' IC/96/17, hep-th/9601376.

\bibitem{kuz} V.A. Kuzmin, M.E. Shaposhnikov and I.I. Tkachev Phys. Rev.
8 (1990) 71.

\bibitem{fuj} Y. Fujimoto and S. Sakakibara, 
Phys. Lett.  B151 (1985) 260;
E. Manesis and S. Sakakibara, Phys. Lett. B157 (1985) 287;
G.A. Hajj and P.N. Stevenson, Phys. Rev. D37 (1988) 413;
K.G.Klimenko, Z.Phys. C43 (1989) 581; Theor. Math. Phys. 80 (1989) 929;
M.P.Grabowski, Z.Phys. C48 (1990), 505.

\bibitem{bim} G. Bimonte and G. Lozano, 
Phys. Lett. B366 (1996) 248; Nucl. Phys. B460 (1996), 155.

\bibitem{ame} G. Amelino-Camelia, 
``On the CJT Formalism in Multi-Field Theories'', hep-th/9603135..

\bibitem{corn} J.M.Cornwall, R.Jackiw and E.Tamboulis, Phys.Rev. D10 
(1974) 2428.

\bibitem{amepi} G.Amelino-Camelia and S-Y.Pi, Phys.Rev. D47 (1993) 2356; G.Amelino-Camelia, Phys.Rev. D49 (1994) 2740.

\bibitem{wet} N. Tetradis and C. Wetterich, Nucl. Phys. B398 (1993) 659;
Int. J. Mod. Phys. A9 (1994), 4029; \\
S. Bornholdt, N. Tetradis and C. Wetterich, ``High Temperature
Phase Transition in Two-Scalar Theories'', OUTP-95-02 P, 
hep-ph/9503282;


\bibitem{roos} T. Ross, ``Wilson Renormalization Group Study 
of Inverse Symmetry Breaking", CLNS 95/1373, hep-th/9511073

\bibitem{wip} Y. Fujimoto, A. Wipf and H. Yoneyama, Phys. Rev. D38 (1988),
2625.

\bibitem{ora} L. O'Raifeartaigh, A. Wipf and H. Yoneyama, Nucl. Phys.
B271 (1986), 653.

\bibitem{KY} C. King and L.G. Yaffe, Comm. Math. Phys. 108
(1987), 423.

\bibitem{roe} G. Roepstorff, {\it Path Integral Approach
to Quantum Physics} (Springer-Verlag, 1994).





\end{thebibliography}
\end{document}